# Dynamically controlling the emission of single excitons in photonic crystal cavities

Francesco Pagliano[1*], YongJin Cho[1], Tian Xia[1], Frank van Otten[1], Robert Johne[1,2] and Andrea Fiore[1]

[1] *COBRA Research Institute, Eindhoven University of Technology, 5600 MB Eindhoven, The Netherlands*

[2] *Max-Planck-Institute for the Physics of Complex Systems, Nöthnitzer Straße 38, 01187 Dresden Germany*

*E-mail: f.m.pagliano@tue.nl

**Single excitons in semiconductor microcavities represent a solid-state and scalable platform for cavity quantum electrodynamics (c-QED), potentially enabling an interface between flying (photon) and static (exciton) quantum bits in future quantum networks. While both single-photon emission and the strong coupling regime have been demonstrated, further progress has been hampered by the inability to control the coherent evolution of the c-QED system in real time, as needed to produce and harness charge-photon entanglement. Here, using the ultrafast electrical tuning of the exciton energy in a photonic crystal (PhC) diode, we demonstrate the dynamic control of the coupling of a single exciton to a PhC cavity mode on a sub-ns timescale, faster than the natural lifetime of the exciton, for the first time. This opens the way to the control of single-photon waveforms, as needed for quantum interfaces, and to the real-time control of solid-state c-QED systems.**

An optical cavity modifies the electromagnetic environment of an emitter and thereby its radiative recombination dynamics. In the weak-coupling (WC) regime, i.e. when the cavity loss rate exceeds the emitter-cavity coupling rate, the increased local optical density of states produces a change in the spontaneous emission rate [1]. This enables the efficient funneling of spontaneously emitted photons in a given mode, and can be used to realize efficient single-photon sources [2][3][4]. In the strong coupling (SC) regime, the emission becomes reversible and coherent Rabi oscillations of the emitter and cavity populations, resulting in a Rabi doublet, are observed [5][6][7]. A dynamic control of the emitter-field interaction is needed to take full advantage of c-QED processes. The real-time control of Rabi oscillations would enable the creation of entanglement between emitter and photon, a fundamental resource in the context of quantum communications [8]. In the WC regime, the ultrafast control allows for the generation of time-symmetric photon wavepackets, enabling the coherent transfer of quantum states between light and matter, a prerequisite for efficient quantum interfaces [9] and spin-memories in QDs [10].

In atomic c-QED system at microwave frequencies, this control is achieved by changing the interaction time [11], while at optical frequencies it is realized by adiabatic passage techniques, which enable shaping a photon waveform on 100 ns timescales [1][2][9][12]. The catch and release of single photons [13] as well as the photon waveform shaping [14] have been recently demonstrated at microwave frequencies using superconducting circuit quantum electrodynamics. Semiconductor c-QED systems, based for example on quantum dots (QDs) in optical microcavities, present an evident potential in terms of integration and scalability. While both single-photon emission [15][16] and the strong coupling regime [5][6] have been demonstrated, the progress in their application to quantum information has been strongly limited by the absence of fast control methods, which would enable the generation of symmetric photons [17][18] as well as the implementation of quantum gates [19][20] and of entangled photon-exciton states. While a number of approaches have been proposed for the control of the exciton-cavity interaction [21-26], none of them has been applied to control c-QED on the ultrafast timescales typical of solid-state systems (10s ps in the SC regime and 100s ps in the WC regime). A method for the all-optical ultrafast control of quality factor and mode volume has been very recently proposed by our group [27], but its application has so far remained limited to ensemble of QDs, eluding the most interesting single-emitter, single-photon regime. Here we report the first demonstration of the *dynamic* control of the radiative interaction of a *single* solid-state emitter with an optical cavity, on a timescale faster than its natural decay (~ns). This is obtained by placing the QD in a photonic crystal cavity integrated with a diode designed to enable a >1 GHz modulation frequency. By the fast tuning of the exciton energy via the quantum confined Stark effect (QCSE), we electrically change the QD-cavity coupling within ~ 300 ps, which is sufficient to reshape the temporal waveform of the photon emitted in the WC regime. Further optimization of the structure will open the way to the control of SC c-QED systems.

## Results

**Device characterization.** In order to be able to create a vertical electric field varying on a sub-ns timescale, we integrate self-assembled QDs inside a GaAs micro-diode with vertical geometry where the top (bottom) part of the 400 nm-thick membrane region is p- (n-)doped. The vertical geometry, as compared to the lateral p-n junctions demonstrated in Ref. [28][29], is essential to provide a large QCSE, since the permanent dipole moment in self-assembled QDs has a vertical orientation, which makes it more sensitive to vertical fields rather than later ones [30]. A small capacitance $C$ ~ 0.3pF and a total series resistance below 200 Ohm were obtained using an optimized design and fabrication (see Methods). A diode (Figure 1) with a mesa diameter of 16 µm and a modified L3 PhCC featuring a y-polarized mode centered around 978 meV at T=10K with a quality factor $Q$ around 10,500 (Figure 1d) is considered in the following. In Figure 1b the 3 dB electrical bandwidth of this diode is shown to be 2 GHz (blue circles). For smaller diodes higher 3 dB frequencies of 4.2 GHz were observed (red circles) but with reduced Q. Micro photoluminescence (mPL) spectra (Fig. 2a) show a clear Stark tuning of the QD excitonic lines as a function of bias voltage V. From the field dependence of the excitonic $QD_1$ line in Figure 2a, which shows Purcell enhancement when tuned inside the cavity mode at V=-1,200 mV, we estimate a permanent dipole moment *p*=(-0.04 ±0.02) e nm and polarizability *β*=(-3.1 ± 0.2) · $10^3$ e nm

kV$^{-1}$ cm, in agreement with similar QD structures [31]. Time-resolved experiments (Fig. 2b) provide the decay time for QD$_1$ out of resonance (400 mV, blue triangles), $\tau^{QD1}_{off}$ = (3.1±0.2) ns, and in resonance (-1,200 mV, red circles), $\tau^{QD1}_{on}$ =(0.27±0.03) ns. The slower component of the in-resonance bi-exponential decay, $\tau_{BG}$ =(2.5±0.2) ns, can be attributed to the nonradiative recombination rate of dark excitons [32]. The emission rate into the mode can be derived correcting for the emission rate into leaky modes, providing a Purcell factor of $F_P = \tau^{BULK} \cdot (1/\tau^{QD1}_{on} - 1/\tau^{QD1}_{off})$ = (2.9 ± 0.6), using the reference time $\tau^{BULK}$ =(0.85±0.10) ns measured on the QD ensemble outside the PhC (green squares). The fraction of photons emitted into the cavity mode $\beta = 1-(\tau^{QD1}_{on}/\tau^{QD1}_{off})$, is calculated as $\beta_{QD1}$= 91% for QD$_1$, comparable with previous reports for L3 cavities [33]. Even higher beta factors may be achieved by positioning the QD at an anti-node of the cavity field by site-controlled growth [34], thus enhancing the Purcell effect and improving the funneling of the emitted photon into the desired mode.

**Ultrafast modulation of the exciton energy.** In order to infer information about the effective modulation of the electric field at the QD position, we study the time-integrated mPL under a square-wave modulation of the applied voltage, as shown in Fig. 3b for a modulation amplitude of 240 mV and frequency varying from 15 MHz to 3 GHz. As sketched in Figure 3a, at low frequencies the voltage and correspondingly the exciton energy have sharp transitions between levels A (-1210mV) and B (-1450mV), resulting in two clear peaks (labeled as A$_i$, B$_i$ in Fig. 3b) for each QD line. The average bias is chosen so that the QD$_1$ line (yellow curves) is in resonance with the cavity in one half of the period (A$_1$). For modulation frequencies above 200 MHz the field across the QDs is affected by the RC constant of the circuit and the two lines get closer to their mean central value. However, the fact that the two peaks are distinct and separated by more than the cavity linewidth at frequencies exceeding 2 GHz demonstrates the possibility to control the QD-cavity coupling on sub-ns timescales.

**Dynamic control of the cQED interaction.** We then investigated the dynamic modulation of the photon waveform by time-resolved PL experiments (TRPL). For this purpose we used a spectral filter to select the emission from the QD$_1$ line and we chose the square-wave bias in order to modulate QD$_1$ in and out of the cavity resonance, but inside the filter window. Figure 3c shows the low power (~ 150 nW) mPL spectra taken at the two modulation voltages of *V*=-1210mV (in resonance, black curve) and *V*=-1260 mV (out of resonance, red curve). The background contribution to the mode feeding [35] [36] is negligible at this low pumping power, as shown by the very low PL intensity observed at the cavity resonance when the QD is detuned by ~ 1 meV (*V*=-1450 mV, blue line in Fig. 3c). Depending on the delay between the laser and the electrical pulses, we observe different scenarios as shown in Fig. 4a-c. From each measured data (black circles), we subtracted the slow contribution due to the dark exciton (red line), as estimated from a bi-exponential fit, to best compare the relevant bright exciton dynamics with our cQED simulation (green lines), based on a master equation formalism (see Methods). During the time interval of Figure 4a QD$_1$ is fixed at its higher energy state showing the natural on-resonance exponential decay. In Fig 4b the QD is initially detuned from the cavity and is brought in resonance at $\Delta t$ ~ 0.25 ns by the electric pulse, thus increasing the spontaneous emission rate. Changing the delay allows controlling the timing of the modulation and gives us the sub-ns temporal control over the modulated cavity output, as shown in Fig. 4c for a larger $\Delta t$ ~ 0.47 ns delay. In this case, the modulated cavity signal is lower since the QD is brought in resonance with the cavity at a later time when its population has already partially decayed. The maximum modulation with respect to the unperturbed mono-exponential decay is a factor of 1.9 (at a time t=1.18 ns in Fig. 4c), and is limited by the maximum QD-cavity detuning of ~ 0.13 meV in this experiment (as seen in the limited ratio of 1.65 in the integrated PL peaks in resonance and out of resonance in Fig. 3(c)). The measurements show an excellent agreement with the c-QED simulations (green lines), taking into account the energy level modulation (orange lines) expected from the RC constant of the circuit, and the QD and cavity parameters extracted from the measurements in Fig. 1 and 2 (see Methods).

## Discussion

The present work clearly demonstrates the dynamic control of the single exciton spontaneous emission within about 300 ps (time between 10% and 90% of the total modulation amplitude in Fig. 4). An ultrafast modulation of the c-QED evolution is achieved when the emitter is dynamically tuned with respect to the cavity resonance. The dynamic control in our experiment is limited by the relatively high QD density, which forces us to use a narrow spectral filter and therefore a limited tuning range, so that the QD emission out of resonance is not negligible. The electrical bandwidth of our PhCC-diode would indeed be sufficient to control the rise time of the photon and achieve a complete symmetrization of its waveform, as estimated in the simulated dynamics of Figure 4d, where the same c-QED parameters as our experiment and a larger tuning range have been assumed. We conclude that by combining our ultrafast diode structure with site-controlled QDs [34], a nearly perfect control of the waveform of emitted single-photons can be achieved. Such symmetric photon waveform, in combination to phase modulation to compensate the chirp induced by the tuning [37], would enable efficient quantum state transfer in quantum networks with solid-state nodes [8][10]. Additionally, by increasing the cavity Q factor up to 60,000, the maximum value allowed by the absorption in the p-doped layer, the control of the Rabi oscillation would be possible for coupling constants g close to the present value, making the long-awaited control of solid-state c-QED processes possible and enabling the generation and control of charge-photon entanglement.

## Methods

**Device Fabrication.** Self assembled InAs QDs were epitaxially grown on GaAs(001) in a Molecular Beam Epitaxy (MBE) reactor with Stranski–Krastanov method, with a density of few tens per $\mu m^2$, an ensemble peak wavelength at cryogenic temperature of about 1270nm and a mean single-exciton linewidth of about 100 $\mu eV$, using the low-growth rate procedure introduced in [38]. They are integrated in the middle of a GaAs intrinsic region that is embedded between two uniformly doped layers (80 nm-thick top GaAs doped p=$1.5 \times 10^{18} cm^{-3}$ and 80 nm-thick bottom GaAs doped n=$2 \times 10^{18} cm^{-3}$). The total thickness (400 nm) of the membrane is chosen as a compromise between low diode capacitance, low sheet resistance of the contact layers and high quality factor. The diode structure is grown on top of an undoped $Al_{0.8}Ga_{0.2}As$ sacrificial layer. Diode mesas with a diameter of 16 $\mu m$, have been defined using optical lithography and wet etched with a calibrated CitricAcid:$H_2O$ solution. On the bottom n-doped layer a low-resistance multi-layer Ge/Ni/Au Ohmic contact is thermally evaporated and annealed at 415 C to preserve the quality and the emission wavelength of the QDs. The $Si_3N_4$ insulation layer is deposited and reopened selectively with a BHF solution in the regions of the mesas and of the n-contacts. A second $Si_3N_4$ hard mask of 380nm is deposited for PhCC fabrication. The PhCC patterns are then defined by exposing a ZEP-520A resist with an electron beam lithography system using proximity effect correction. The PhCC holes are then transferred to the SiN hard-mask by reactive ion etching and finally to the diode membrane using a $SiCl_4$:Ar recipe. The $Al_{0.8}Ga_{0.2}As$ sacrificial layer is undercut to a lateral extent of 2 µm around the PhCC pattern in 10% hydrofluoric acid. The SiN hard-mask is finally removed by a low power $CF_4$:O2 plasma that preserves the underlying SiN insulation without damaging the GaAs surface. The top Zn/Au contact is then evaporated. This fabrication technique allows us to achieve a PhCC with quality factor around $10^4$ and a GHz electrical bandwidth in the same device. The cavity resonance wavelength is lithographically tuned in order to match the ground state emission wavelength of the QD ensemble at *T*=10 K. Several modified L3 cavities, one for each device, are fabricated with lattice parameter *a* ranging from 290 nm to 320 nm and constant ratio *r/a*=0.32, with *r* the non modified holes radius. To enhance the quality factor of the cavity a modified L3 designed is used shifting the first two holes adjacent to the cavity by about 0.15 *a* and reducing their diameter by 20% [39]. The ideal $Q_{RAD}$ factor for this cavity, calculated from FEM simulations, is around 27,000. From the SEM analysis we have estimated a standard deviation of $\sigma$~ 2.3 nm in the holes radii dimensions and positions, which may be partly responsible for the lower experimental Q value of 10,500. Moreover an absorption of $\alpha$~$10^2$ cm$^{-1}$ in the p-doped GaAs layer set a limit of  ~ 60,000 for the real $Q_{TOT}$ = $Q_{RAD}*Q_{ABS}/(Q_{RAD}+Q_{ABS})$.

**Electro-optical characterization.** The sample is tested electrically and optically in a cryogenic probe-station setup where a RF micro-probe is used to bias the diodes one by one. A high numerical aperture objective is used to excite the devices and collect the emitted photons. The radiative signal is coupled to a single mode fiber for spectral characterization and time-resolved experiments. The diodes IV curves have a turn-on voltage around 1.1 V at *T*=10 K and a reverse bias current of order of few nA with breakdown voltages that range from -5 V to -9 V for different devices. To estimate the impedances of the device and its 3 dB frequency we use a network analyzer to measure the scattering parameter $S_{11}$= *(Z- $Z_0$)/(Z+ $Z_0$)* , $Z_0$ being the 50 Ohm reference resistance and *Z=R-j/($2\pi f$ C)* the frequency dependent total impedance of the circuit. From the amplitude and the phase of $S_{11}$ both R and C can be estimated. A record 3 dB frequency of up to 4.2 GHz was obtained for optimized devices with a mesa diameter $\phi$=12 µm (red circles of Figure 1b). Although the larger diodes ($\phi$=16 µm) show a lower electrical bandwidth, they represent a better compromise between speed and PhCC quality factor.

The time-resolved PL data in Fig. 4 is obtained in an experiment where the photons emitted by $QD_1$ are collected by the confocal microscope, selected with a spectral filter (full-width half-maximum FWHM=0.65 nm) at the $QD_1$ energy and sent through a fiber to a Superconducting Single Photon Detector (SSPD), whose output signal is correlated with the trigger from the pulsed laser controller. To show the possibility to dynamically tune the QD at a specific time, we implemented a sub-ns synchronization between the optical excitation pulse and the rising edge of the square wave bias. The first channel of our fast bias generator provides the external trigger to the pulsed laser diode at 970 nm while the second channel is used to generate the electrical function for the dynamic Stark tuning of $QD_1$. In this configuration the two channels share the same trigger frequency of 60 MHz and a variable delay can be set between the optical and the electrical pulse while the filtered signal is collected from the device and sent to the SSPD every 16.67 ns. While the optical and the electrical paths are kept fixed, the relative delay between the laser and the voltage pulses is changed by varying the internal delay between the first and the second channel of the bias generator with sub-ns accuracy.

**c-QED simulations.** To theoretically describe the experimental situation, the QD is modeled as a three-level quantum emitter coupled to the field of an optical microcavity. In this scheme the ground state of the QD (level 1) is incoherently pumped at time $t_0$ (operator $\sigma_{13}$) to the higher energy state (level 3) which then relaxes incoherently to the cavity coupled excitonic state of the QD (level 2) through the lowering operator $\sigma_{23}$. The dynamics of the system is simulated by the temporal evolution of the density matrix. The system Hamiltonian describing the coherent QD-cavity interaction in a frame rotating at $\omega_{FRAME}$ reads

$$H = \Delta_{CAV}\hat{a}^\dagger\hat{a} + \Delta_{QD}(t)\hat{\sigma}_{12}^\dagger\hat{\sigma}_{12} + i\frac{\hbar}{2}(a\,\sigma_{12} - \hat{\sigma}_{12}^\dagger\hat{a})$$

where $a$ ($a^\dagger$) is the bosonic annihilation (creation) operator for the cavity field, $\sigma_{12}$ is the lowering operator that couples the QD exciton state to the cavity mode transferring the quantum emitter to its ground state and $\Omega=2g$ is the Rabi frequency of the coupled system. The cavity frequency is fixed and the detunings are given by $\Delta_i = \omega_i - \omega_{FRAME}$ with $\Delta_{CAV} = 0$. The time-dependent QD frequency $\omega_{QD}(t)$ follows the piecewise function (orange lines of Figure 4) that best approximates the steady state response of the diode to a 50% duty cycle square wave modulation well below its cut-off frequency and is written as

$$\omega_{QD}(t) = \omega_A + (\omega_B - \omega_A) \times \begin{cases} C\,e^{\left(-\frac{T}{2\tau_{RC}}\right)}, & t < \delta \\ 1 - C\,e^{\left(\frac{\delta-t}{\tau_{RC}}\right)}, & \delta < t < \delta + \frac{T}{2} \\ C\,e^{\left(\frac{\delta+\frac{T}{2}-t}{\tau_{RC}}\right)}, & t > \delta + \frac{T}{2} \end{cases}$$

where $C = 1/(1+ e^{-T/2\tau_{RC}})$ and the temporal constant of the circuit $\tau_{RC}$ is set to 140 ps to reproduce the behavior of the diode with an experimental electro-optical bandwidth around 1.2 GHz (estimated from Figure 3b). The initial (final) QD frequency is denoted by $\omega_A$ ($\omega_B$), $\delta$ is the delay at which the tuning is occurring and T = 16.67 ns the period of the electrical signal corresponding to the experimental rate of 60 MHz. The modulation is simulated between the two static energies $E_A$ = 978.02 meV and $E_B$ = 977.89 meV.

Including the incoherent pumping, the relaxation of the higher energy level to the bright exciton state and the system losses, the Master equation for the complete dynamics reads

$$\dot{\rho} = -\frac{i}{\hbar}[H,\rho] + \kappa_{CAV}\left(\hat{a}\rho\hat{a}^\dagger - \frac{1}{2}\hat{a}^\dagger\hat{a}\rho - \frac{1}{2}\rho\hat{a}^\dagger\hat{a}\right) + \gamma_{QD}\left(\hat{\sigma}_{12}\rho\hat{\sigma}_{12}^\dagger - \frac{1}{2}\hat{\sigma}_{12}^\dagger\hat{\sigma}_{12}\rho - \frac{1}{2}\rho\hat{\sigma}_{12}^\dagger\hat{\sigma}_{12}\right)$$
$$+ \gamma_{relax.}\left(\hat{\sigma}_{23}\rho\hat{\sigma}_{23}^\dagger - \frac{1}{2}\hat{\sigma}_{23}^\dagger\hat{\sigma}_{23}\rho - \frac{1}{2}\rho\hat{\sigma}_{23}^\dagger\hat{\sigma}_{23}\right) + \Phi_{PUMP}\left(\hat{\sigma}_{13}^\dagger\rho\hat{\sigma}_{13} - \frac{1}{2}\hat{\sigma}_{13}\hat{\sigma}_{13}^\dagger\rho - \frac{1}{2}\rho\hat{\sigma}_{13}\hat{\sigma}_{13}^\dagger\right)$$

where $\kappa_{CAV}$ is the cavity loss rate, $\gamma_{QD}$ is the exciton radiative decay rate into the leaky modes of the photonic crystal, $\gamma_{relax.}$ is the fast incoherent relaxation rate from the pumped high energy level to the exciton state and $\Phi_{PUMP}$ is the incoherent pump term. We solve the Master equation numerically. The cavity damping rate, the QD decay rate into leaky modes and the QD-cavity coupling rate have been extracted from the data in Figs. 1 and 2 to be $\kappa_{CAV}/2\pi$ =22 GHz, $\gamma_{QD}/2\pi$=0.036 GHz, $g/2\pi$ =5.4 GHz, respectively. The precise electric field dependence of the QD energy and the electrical delay has been optimized in order to match the experimental data and is shown in Fig. 4. The emitted photon intensity from the QD-cavity system during the QD tuning is calculated as the mean value of the cavity loss rate using the relation $\langle \kappa_{CAV}\hat{a}^\dagger\hat{a}\rangle = Tr[\kappa_{CAV}\hat{a}^\dagger\hat{a}\rho]$.

## Acknowledgements

This work is part of the research programme of the Foundation for Fundamental Research on Matter (FOM), which is financially supported by the Netherlands Organization for Scientific Research (NWO), and is also supported by the Dutch Technology Foundation STW, applied science division of NWO, the Technology Program of the Ministry of Economic Affairs under Project No. 10380. The nanofabrication work has been performed in the NanoLab@TU/e clean-room.  We acknowledge B.T de Vries, E.

Smalbrugge and E. Geluk for the support with the fabrication facilities and T.B. Hoang for the help in the experiments.## Author contributions

A.F. proposed the experiment and led the project. Y.C., T.X. and F.v.O. optimized and performed the sample growth, F.P. designed and fabricated the devices and performed the measurements, F.P. and R.J. performed the simulations. F.P., A.F. and R.J. discussed the results and prepared the manuscript.

## Competing financial interests

The authors declare that they have no competing financial interests.

# Figures

## Figure 1: Ultrafast Photonic Crystal Cavity Diode

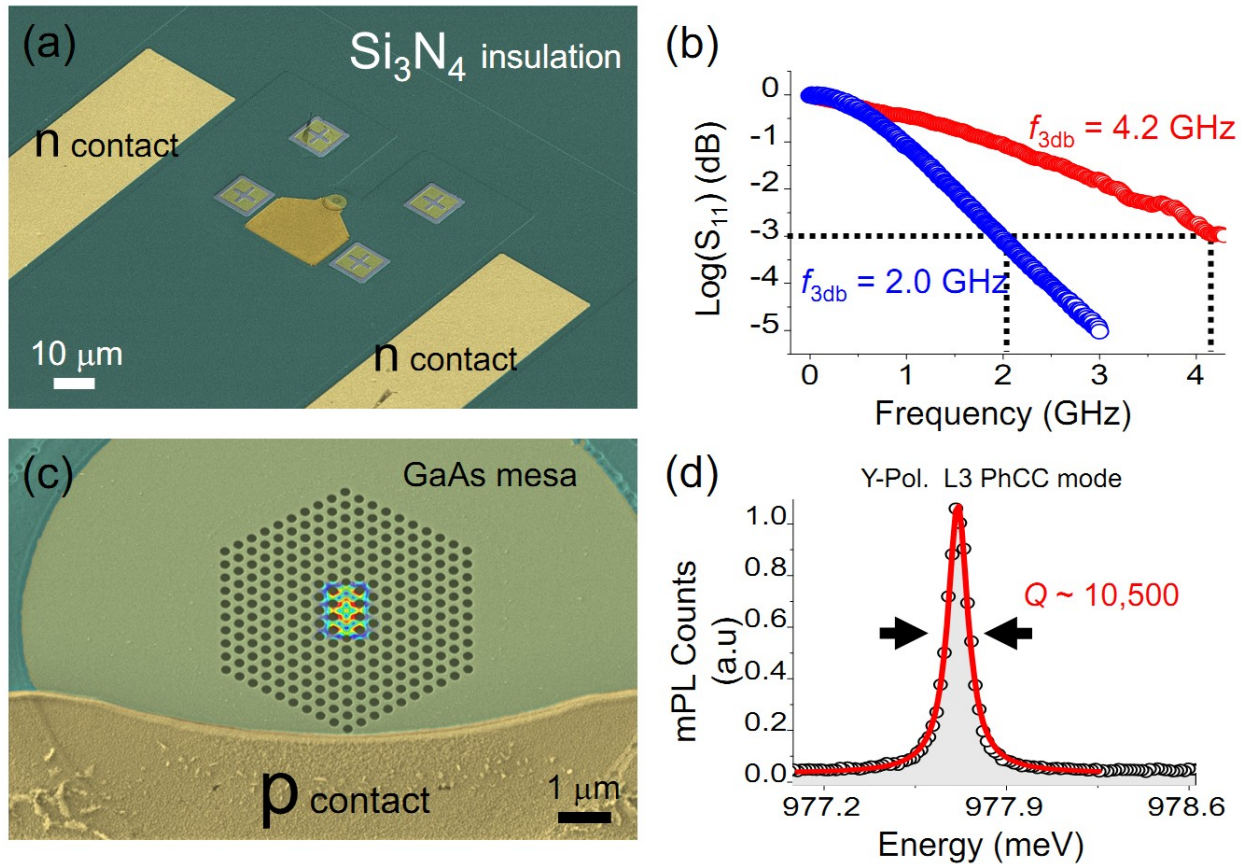

Scanning electron microscope images of the photonic crystal cavity diode (**a**) and enlarged view of the L3 modified cavity with the electric field $|\mathbf{E}|^2$ distribution (over-imposed) of the $Y_1$ mode calculated from 3D FEM simulation (**c**). (**b**) Network analyzer measurement of the $S_{11}$ one-port scattering parameter showing the diode bandwidth of 2 GHz (blue circles) compared to the data of a smaller device with better electrical performance (red circles). (**d**) High resolution spectrum of the PhCC L3 cavity $Y_1$ mode with Lorentzian fit (red curve) and estimated Q-factor of 10,500 ($k/2\pi \sim$ 22 GHz).

**Figure 2:** Static Stark tuning and spontaneous emission control

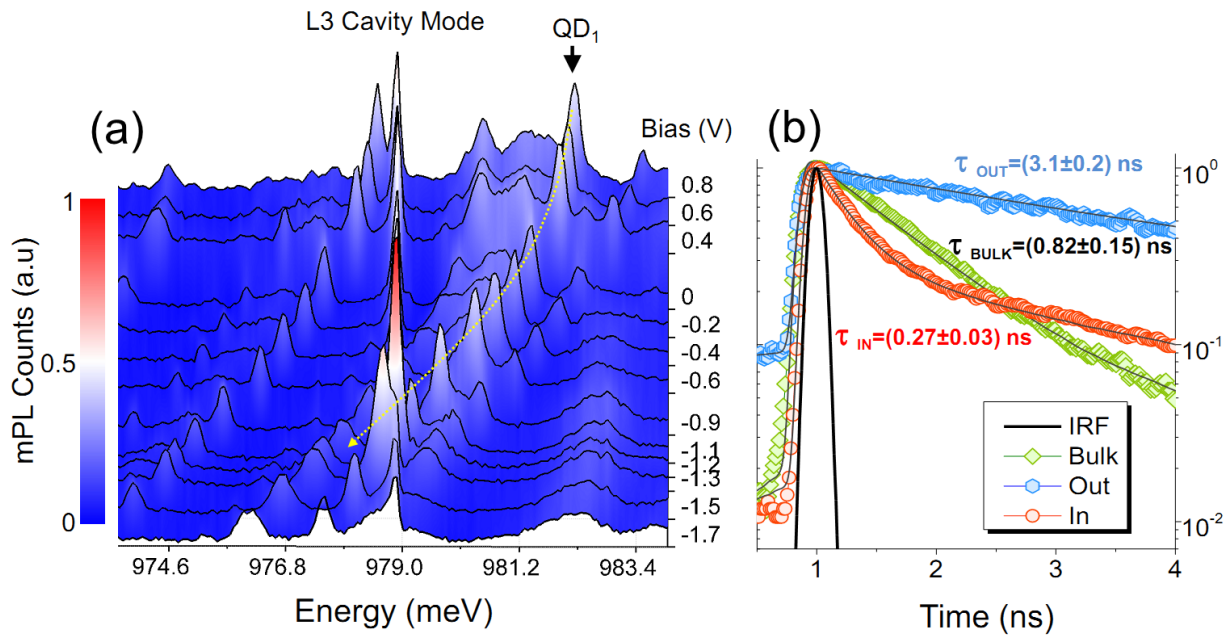

(a) Low temperature mPL spectra for 800 nW excitation power for different DC voltages applied to the diode. A weakly coupled single exciton line labeled $QD_1$ is enhanced when tuned across the L3 cavity mode. The yellow line is a guide to the eye. (**b**) TRPL decay curves of $QD_1$ out of (light blue hexagons) and in (red circles) resonance with the cavity mode for $T$=10 K. The QD ensemble decay in the bulk is measured as reference (green curve) and the decay time constants are calculated from single and double exponential fits (grey lines) including the convolution with the IRF of the detector (black line).

**Figure 3:** Squarewave bias modulation of exciton energy

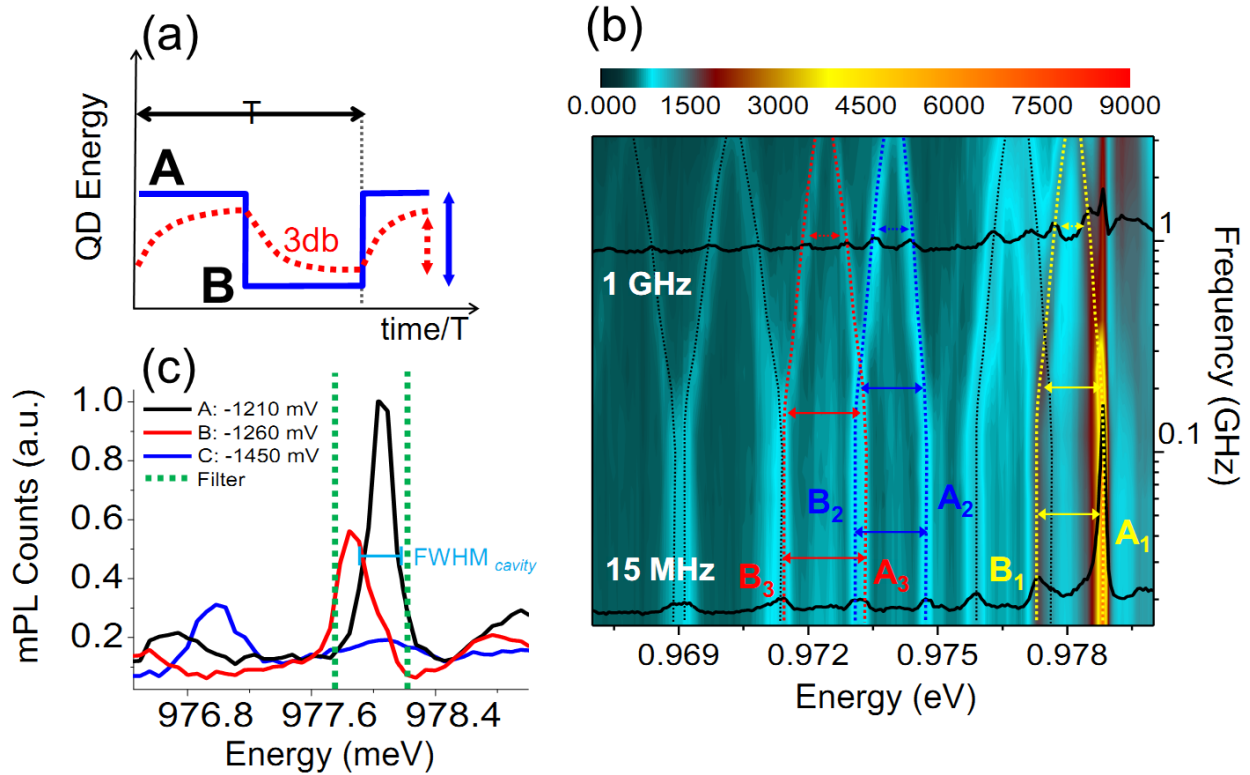

(**a**) Schematics for the low- (blue curve) and the high- (dotted red line) frequency exciton modulation between energies A and B due to the square wave bias in the center of the diode. (**b**) mPL data collected from the QDs inside the cavity and dynamically modulated by the square wave bias between -1210mV and -1450mV at different frequencies up to 3GHz. Two lines outside of the cavity mode spectral region are modulated between the energy states $A_2$-$B_2$ (bue curves and arrows) and $A_3$-$B_3$ (red curves and arrow) while the coupled $QD_1$ is tuned in($A_1$) and out($B_1$) of resonance with the cavity mode and enhanced. (**c**) Low-excitation spectra (~150 nW) for the $QD_1$ at -1210 mV (black), -1260 mV (red) and -1450 mV (blue). The green dotted lines show the filter window (FWHM=0.65 nm) in which the fast modulated cavity signal is collected and sent to the SSPD for the measurements of Figure 4.

**Figure 4:** Ultrafast electrical control of QD$_1$-cavity coupling

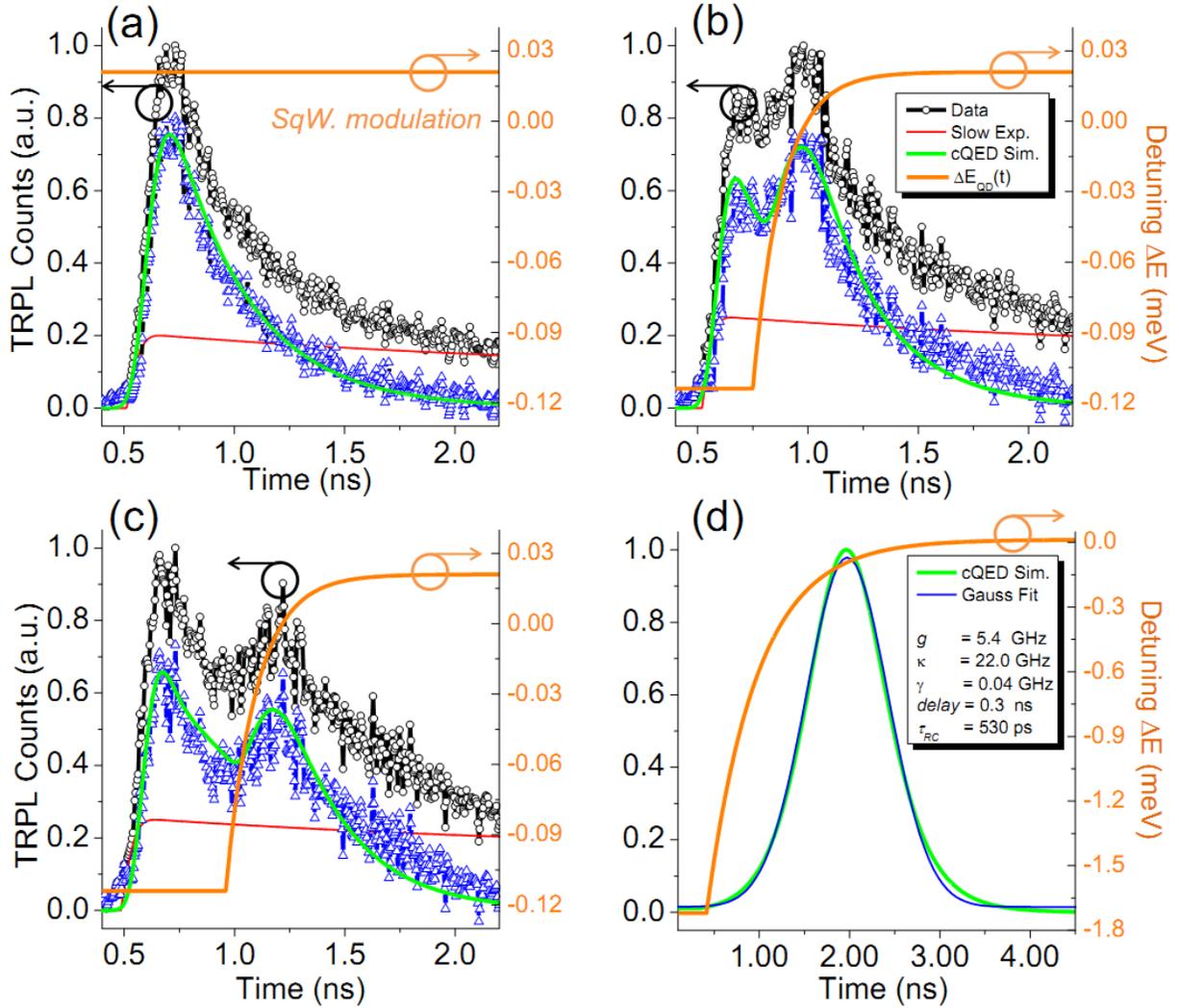

Time-resolved mPL experiments under quasi-resonant excitation. A pulsed laser at 970 nm with a pulse width of ~70 ps at a repetition rate of 60 MHz is used. The fast modulation of the normalized photon counts (black circles) collected from the cavity center during the dynamic tuning of QD$_1$ energy in (-1210 mV) and out (-1260 mV) of resonance with the cavity mode is shown for different delays between the laser excitation and the square wave electric signal. The fast component of this filtered signal shows a good agreement with the theoretical simulations (green curves) not including the slow exponential term (red lines). The experimental delays can be more precisely estimated from the simulation and are found to be $\Delta t > 3$ ns (**a**), $\Delta t = (0.25 \pm 0.03)$ ns (**b**) and $\Delta t = (0.47 \pm 0.01)$ ns (**c**). (**d**) Simulation of the photon symmetrization and Gaussian fit (blue line) with the same cQED parameters, a larger tuning range and a square wave bias filtered with a RC time constant of 530 ps.